\documentclass[11pt]{article}
\usepackage{amssymb}
\usepackage{amsmath}
\usepackage{amscd}
\usepackage{latexsym}
\usepackage{bbm}

\catcode`@=11
\makeatletter\renewcommand{\section}{\@startsection
{section}{1}{\z@}{-3.5ex plus -1ex minus
    -.2ex}{2.3ex plus .2ex}{\large\bf }}

\makeatletter\renewcommand{\subsection}{\@startsection{subsection}{2}{\z@}{-3.25ex
plus -1ex minus
   -.2ex}{1.5ex plus .2ex}{\bf }}

\numberwithin{equation}{section}
\newcounter{saveeqn}

\catcode`@=12

\def\a{\alpha}
\def\b{\beta}
\def\ga{\gamma}
\def\la{\lambda}

\def\de{\delta}
\def\eps{\epsilon}
\def\ve{\varepsilon}
\def\vk{\varkappa}

\def\p{\phi}

\newcommand{\C}{\mathbb C}
\newcommand{\R}{\mathbb R}

\newcommand{\Gcal}{{\cal G}}
\newcommand{\Acal}{{\cal A}}

\newcommand{\Mcal}{{\cal M}}
\newcommand{\Fcal}{{\cal F}}
\newcommand{\Ncal}{{\cal N}}

\newcommand{\gfrak}{{\mathfrak g}}

\newcommand{\mh}{{\hat{\smash{\mu}}}}
\newcommand{\nh}{{\hat{\smash{\nu}}}}

\def\1{{\bar 1}}

\def\tr{\textrm{tr}}
\def\diff{\textrm{d}}

\def\sfrac#1#2{{\textstyle\frac{#1}{#2}}}
\def\+{\dagger}
\def\={\ =\ }

\def\and{\quad\textrm{and}\quad}
\def\with{\quad\textrm{with}\quad}

\begin{document}

\begin{titlepage}
\setcounter{page}{0}
\begin{flushright}
.
\end{flushright}

\hspace{2.0cm}

\begin{center}

{\Large\bf
Scattering of instantons, monopoles and vortices\\[2mm] in higher dimensions
}

\vspace{8mm}

{\large Tatiana A. Ivanova
}\\[8mm]

\noindent{\em Bogoliubov Laboratory of Theoretical Physics, JINR} \\
{\em  141980 Dubna, Moscow Region, Russia}\\[2mm]
Email: {\tt ita@theor.jinr.ru}\\[6mm]

\vspace{10mm}

\begin{abstract}
\noindent
We consider Yang-Mills theory on manifolds $\R\times X$ with a $d$-dimensional
Riemannian manifold $X$ of special holonomy admitting gauge instanton equations. Instantons
are considered as particle-like solutions in $d+1$ dimensions whose static configurations
are concentrated on $X$. We study how they evolve in time when considered as solutions of the Yang-Mills
equations on $\R\times X$ with moduli depending on time $t\in\R$. It is shown that in the adiabatic
limit, when the metric in the $X$ direction is scaled down, the classical dynamics of slowly moving instantons
corresponds to a geodesic motion in the moduli  space $\Mcal$ of gauge instantons on $X$. Similar
results about geodesic motion in the moduli space of monopoles and vortices in higher dimensions are briefly discussed.
\end{abstract}

\end{center}
\end{titlepage}

\section{Introduction and summary}

Instantons in four dimensions are nonperturbative Bogomolny-Prasad-Sommerfield (BPS) configurations solving
first-order anti-self-duality equations for gauge fields which imply the full Yang-Mills equations~\cite{1}. If one makes the
assumption that the non-abelian gauge potential is independent of one coordinate on $\R^4$ then the anti-self-duality
equations are reduced to Bogomolny equations on $\R^3$ describing non-abelian monopoles~\cite{2}. Furthermore,
considering anti-self-dual Yang-Mills equations on a four-manifold $\Sigma_2\times S^2$ and imposing SO(3)-equivariance
condition on gauge fields, one obtains
vortex equations on a two-dimensional Riemannian manifold $\Sigma_2$ (see e.g.~\cite{2,3} and references therein).
Vortices, monopoles and instantons are important objects in modern field theories describing nonperturbative
physics~\cite{4}-\cite{7}.

Non-abelian monopoles are also particle-like static solutions of Yang-Mills-Higgs equations in Minkowski space
$\R^{3,1}$~\cite{1}. Vortices can also be obtained as static solutions of Yang-Mills-Higgs equations in $2+1$
dimensions~\cite{4}-\cite{7}. Similarly instantons can be considered as solitons in $4+1$ dimensional Yang-Mills theory. One
can ask about the dynamics of all these solitons which can evolve according to the second-order field equations of
Yang-Mills-Higgs or Yang-Mills theory. In the seminal paper~\cite{8} Manton suggested that in the ``slow motion limit" the
monopole dynamics can be described in terms of geodesics in the moduli space of static multi-monopole solutions. These
geodesics are defined via a metric on the multi-monopole moduli space~\cite{8}. This heuristic approach was extended  to
vortices in  $2+1$ dimensions~\cite{9}, domain walls in $3+1$ dimensions~\cite{9a} and instantons in $4+1$ dimensions (see
e.g.~\cite{10}). Higher derivative corrections to the lowest-order (adiabatic) results were considered in~\cite{10a}. The
Manton approach was rigorously justified both for monopoles and vortices by Stuart~\cite{11}. However, we are not aware about
such a justification for scattering instantons in $4+1$ dimensions. Here we provide a derivation of motion of instantons along
geodesics in the multi-instanton moduli space by using the adiabatic approach. Generalizing~\cite{10}, we will consider this
approach for instantons not only in four but also in higher dimensions.

Instanton equations  on a $d$-dimensional Riemannian manifold $X$ can be introduced  as follows~\cite{12}-\cite{14}.
Suppose there exists  a 4-form $Q$ on $X$. Then there exists a $(d{-}4)$-form $\Sigma:=*Q$, where $*$ is the Hodge
operator on $X$ defined with the help of a metric $g$ on $X$.
Let $\Acal$ be a connection on a rank-$k$ vector bundle $E$ over $X$ with the curvature $\Fcal_X =\diff\Acal + \Acal\wedge\Acal$.
For simplicity we choose SU($k$) as our gauge group and therefore both $\Acal$ and  $\Fcal_X$ take values in the Lie
algebra su$(k)$. The generalized anti-self-duality (instanton) equation on the gauge field $\Fcal_X$ is~\cite{14}
\begin{equation}\label{1.1}
 *\Fcal_X + \Sigma\wedge\Fcal_X \=0\ .
\end{equation}
For $d>4$ these equations can be defined on manifolds $X$ with {\it special holonomy}, i.e.\ such that the holonomy
group $G^h$
of the Levi-Civita connection on the tangent bundle $TX$ is a subgroup in SO$(d)$. Solutions of (\ref{1.1}) satisfy the
Yang-Mills equations
\begin{equation}\label{1.2}
D_\mu\Fcal^{\mu\nu}_X:=\frac{1}{\sqrt{\det g}}\,\partial_\mu(\sqrt{\det g}\,\Fcal^{\mu\nu}_X) + [\Acal_\mu ,
\Fcal^{\mu\nu}_X]\=0\ ,
 \end{equation}
where the derivatives $\partial_\mu :=\partial/\partial x^\mu$ are taken with respect to local coordinates
$x^\mu$ on $X$ and $g=g_{\mu\nu}\, \diff x^\mu\diff x^\nu , \mu ,\nu , ...=1,...,d$.
The instanton equations  are also well defined on manifolds $X$ with non-integrable $G^h$-{\it structures}, i.e.\
when $\diff\Sigma\ne 0$. In this case  (\ref{1.1}) imply the Yang-Mills equations with (3-form) torsion
$T:=*\diff\Sigma$, as is discussed e.g.\ in~\cite{15}-\cite{18}. Such torsionful Yang-Mills equations
naturally appear in heterotic string compactifications with $H$-flux.

We extend the manifold $X$ by the time axis $\R$ and introduce on the Lorentzian manifold $M=\R\times X$ a metric
\begin{equation}\label{1.3}
\hat g_\ve = -\diff t^2 + \ve^2g\ ,
 \end{equation}
where $t=x^0$ is a coordinate on $\R$ and $\ve$ is a real parameter. Denoting by $\{x^\mh\}=\{x^0, x^\mu\}$ local
coordinates on $M=\R\times X$, we introduce the Yang-Mills equations on $M$,
\begin{equation}\label{1.4}
\hat D_\mh\hat\Fcal^{\mh\nh}:=\frac{1}{\sqrt{\det g}}\,\partial_\mh(\sqrt{\det g}\,\hat\Fcal^{\mh\nh}) +
[\hat\Acal_\mh , \hat\Fcal^{\mh\nh}]\=0\ ,
\end{equation}
where we used the fact that $|\det\hat g_\ve|=\ve^{2d}\det g$.

It is not easy to construct non-trivial time-dependent solutions of the Yang-Mills equations (\ref{1.4}). The {\it adiabatic
limit} method, based on Manton's idea, provides a useful and powerful tool for describing such solutions. The adiabatic limit
refers to the geometric process of shrinking the metric  (\ref{1.3}) in the $X$ direction by taking the limit $\ve\to 0$. We
will show that solutions of the Yang-Mills equations (\ref{1.4}) in the limit $\ve\to 0$ for the metric  (\ref{1.3}) converge
to the solutions of one-dimensional sigma-model describing a map from $\R$ into the moduli space of gauge instantons on $X$.
For connections $\Acal$ not depending on one coordinate of $X$ we will get geodesics in the moduli space of (generalized)
monopoles on a $(d-1)$-dimensional submanifold of $X$. Similar reductions to geodesic  in moduli space of (generalized)
vortices on $(d-2)$-dimensional submanifolds of $X$ will also be described.

\section{Moduli space of instantons in $d\ge 4$}

{\bf Moduli space of connections.}\footnote{In this section we follow the discussion of \cite{19}.}
Let $X$ be an oriented smooth manifold of dimension $d$, $G$ a semisimple  compact Lie
group, $\gfrak$ its Lie algebra, $P$ a principal $G$-bundle over $X$, $\Acal$ a connection 1-form on $P$ and
$\Fcal_X =\diff\Acal + \Acal\wedge\Acal$ its curvature. We consider also the bundle of groups Int$P = P\times_G G$
($G$ acts on itself by internal automorphisms: $h\mapsto ghg^{-1},\ h, g\in G$) associated with $P$, the bundle
of Lie algebras Ad$P=P\times_G \gfrak$ and a complex vector bundle $E=P\times_GV$, where $V$ is the space of some
irreducible representation of $G$. All these associated bundles inherit their connection $\Acal$ from $P$. For the
simplicity one can consider $G=\,$SU($k$), $\gfrak =\, $su$(k)$ and $V=\C^k$

We denote by $\mathbb{A}'$ the space of connections on $P$ and by $\Gcal'$ the
infinite-dimensional
group of gauge transformations (automorphisms of $P$ which induce the identity transformation of $X$),
\begin{equation}\label{2.1}
\Acal\ \mapsto\ \Acal^g\=g^{-1}\Acal g + g^{-1}\diff g\ ,
\end{equation}
which can be identified with the space $\Gamma (X,$ Int$P$) of global sections of the bundle Int$P$. Correspondingly,
the infinitesimal action of $\Gcal'$ is defined by global sections  $\chi$ of the bundle Ad$P$,
\begin{equation}\label{2.2}
\Acal\ \mapsto\ \de_\chi\Acal\=\diff\chi + [\Acal , \chi ]\ =:\ D^{}_{\Acal}\chi
\end{equation}
with $\chi\in\, $Lie$\,\Gcal'=\Gamma (X,\, $Ad$P)$.

We restrict ourselves to the subspace $\mathbb{A}\subset\mathbb{A}'$
of irreducible connections and to the subgroup $\Gcal = \Gcal' \,/ Z(\Gcal')$ of $\Gcal'$ which acts freely on $\mathbb{A}$.
Then the {\it moduli space\/} of irreducible connections on $P$ (and on $E$) is defined as the quotient $\mathbb{A}/\Gcal$.
Classes of gauge equivalent connections are points $[\Acal ]$ in $\mathbb{A}/\Gcal$.

Since $\mathbb{A}$ is an affine space, for each $\Acal\in\mathbb{A}$
we have a canonical identification between the tangent space $T_{\Acal} \mathbb{A}$ and the space $\Lambda^1(X,\,$Ad$P$)
of 1-forms on $X$ with values in the vector bundle Ad$P$.
Our $\gfrak =\, $su$(k)$ is a matrix Lie algebra, with the metric
defined by the trace. The metrics on $X$ and on the Lie algebra su$(k)$ induce an inner product
on $\Lambda^1(X,\,$Ad$P$),
\begin{equation}\label{2.3}
 \langle\xi_1, \xi_2\rangle \= \int_X\tr\,(\xi_1\wedge *\xi_2)
 \qquad\textrm{for}\qquad \xi_1, \xi_2\in \Lambda^1(X,\textrm{Ad}P)\ .
\end{equation}
This inner product is transferred to $T_{\Acal} \mathbb{A}$ by the canonical
identification. It is invariant under the $\Gcal$-action on $\mathbb{A}$, whence
we get a metric (\ref{2.3}) on the moduli space $\mathbb{A}/\Gcal$.

\medskip

\noindent {\bf Instanton connections.}
Suppose there exists a $(d{-}4)$-form $\Sigma$ on $X$ which allows us to introduce the instanton
equation
\begin{equation}\label{2.4}
 *\Fcal_X + \Sigma\wedge\Fcal_X \=0
\end{equation}
discussed in Section 1.
We denote by $\Ncal\subset\mathbb{A}$ the space of irreducible connections subject to (\ref{2.4})
on the rank-$k$ complex vector bundle $E\to X$. This space $\Ncal$ of instanton solutions on $X$ is a subspace of the
affine space $\mathbb{A}$, and we define the moduli space $\Mcal$ of instantons as the quotient space
\begin{equation}\label{2.5}
\Mcal \=\Ncal /\Gcal
 \end{equation}
together with a projection
\begin{equation}\label{2.6}
\pi :\ \Ncal \stackrel{\Gcal}\to\Mcal \ .
\end{equation}

According to the bundle structure (\ref{2.6}), at any point $\Acal\in \Ncal$,
the tangent bundle $T_{\Acal}\Ncal\to\Ncal$ splits into the direct sum
\begin{equation}\label{2.7}
T_{\Acal}\Ncal \=\pi^*T_{[\Acal ]}\Mcal\oplus T_{\Acal}\Gcal \ .
\end{equation}
In other words,
\begin{equation}\label{2.7a}
T_{\Acal}\Ncal \ni\ \tilde\xi \= \xi + D_\Acal\chi \qquad\textrm{with}\qquad
\xi\in \pi^*T_{[\Acal ]}\Mcal  \and  D_\Acal\chi\in T_{\Acal}\Gcal\ ,
\end{equation}
where  $\tilde\xi , \xi\in \Lambda^1(X,\,$Ad$P$) and $\chi\in \Lambda^0(X,\,$Ad$P)=\Gamma(X,\,$Ad$P$).
The choice of $\xi$ corresponds to a local fixing of a gauge. We
denote by $\xi_\a$
a local basis of vector fields on $\Mcal$ (sections of the tangent bundle $T\Mcal$) with $\a = 1,\ldots,
\textrm{dim}_\R\Mcal$.
Restricting the metric (\ref{2.3}) on $\mathbb{A}/\Gcal$ to the subspace $\Mcal$
provides a metric ${\mathbb G}=(G_{\a\b})$ on the instanton moduli space,
\begin{equation}\label{2.8}
 G_{\a\b}\=\int_X \tr\,(\xi_\a\wedge *\xi_\b)\ .
\end{equation}
Using this metric on $\Mcal$, we can introduce Christoffel symbols
\begin{equation}\label{2.10}
\Gamma^{\ga}_{\a\b}=\frac{1}{2}\, G^{\ga\vk}\bigl(\partial_\a G_{\b\vk}+\partial_\b G_{\a\vk}-
\partial_{\vk} G_{\a\b}\bigr)\ ,
\end{equation}
where the derivatives $\partial_\a:=\partial/\partial\phi^\a$ are taken with respect to local
coordinates $\phi^\a$ on $\Mcal$ in which ${\mathbb G} =G_{\a\b}\,\diff\phi^\a\, \diff\phi^\b$. One can also introduce
Riemannian tensor, Ricci tensor etc.

\section{Adiabatic limit for the Yang-Mills equations in $d\ge 4$}

\noindent {\bf Splitting of the Yang-Mills equations.} So, we consider the manifold
\begin{equation}\label{3.1}
M=\R\times X
\end{equation}
with a metric
\begin{equation}\label{3.2}
\hat g_\ve = -\diff t^2 + \ve^2g=-\diff t^2 + \ve^2g_{\mu\nu}\diff x^\mu\diff x^\nu\ ,
\end{equation}
and rank-$k$ complex vector bundle $E\to M$ with an su$(k)$-valued connection $\Acal$ as well as
the curvature 2-form
\begin{equation}\label{3.3}
\hat\Fcal = \sfrac12\, \Fcal_{\hat\mu\hat\nu}\diff x^{\hat\mu}\wedge\diff x^{\hat\nu} =
\Fcal_{0\mu}\diff x^0\wedge\diff x^\mu + \sfrac12\, \Fcal_{\mu\nu}\diff x^\mu\wedge\diff x^\nu\ .
\end{equation}
Recall that we assume that the second part in (\ref{3.3}),
\begin{equation}\label{3.4}
\Fcal_X = \sfrac12\, \Fcal_{\mu\nu}\diff x^\mu\wedge\diff x^\nu \ ,
\end{equation}
satisfies the instanton equation (\ref{2.4}) and for the connection $\hat\Acal$ on $E\to \R\times X$
we have
\begin{equation}\label{3.5}
\hat\Acal= \Acal_{\hat\mu}\diff x^{\hat\mu} =\Acal_{0}\diff x^{0}+ \Acal_{\mu}\diff x^{\mu}=\Acal_0\diff t +\Acal
\end{equation}
where $\Acal$ has components only along $X$ but depends on all coordinates $(t,x^\mu )$ on $M$.

We assume that $\Acal$ satisfies the instanton equation (\ref{2.4}) for any $t$ and depend on $t$ only via
moduli $\phi^\a$ (collective coordinates) described in Section 2. On the other hand, the full Yang-Mills equations
(\ref{1.4}) impose restrictions on dynamics of $\phi^\a(t)$. In order to find them we note that for the
metric (\ref{3.2}) we have
\begin{equation}\label{3.6}
\hat\Fcal^{0\mu}= \hat g^{00} \hat g^{\mu\nu}\Fcal_{0\nu}=\ve^{-2}\Fcal^{0\nu}\ ,\quad
\hat\Fcal^{\mu\nu}=\ve^{-4}\Fcal^{\mu\nu}\ ,
\end{equation}
where in $\Fcal^{0\mu}$ and $\Fcal^{\mu\nu}$ indices are raised by $g^{00}$ and $g^{\mu\nu}$. After substitution
of (\ref{3.6}) into (\ref{1.4}) we obtain the equations
\begin{equation}\label{3.7}
D_\mu\Fcal^{\mu 0}\equiv g^{\mu\nu}D_\mu\Fcal_{0\nu}=0\ ,
\end{equation}
\begin{equation}\label{3.8}
g^{\mu\nu}D_0\Fcal_{0\nu} = 0\ ,
\end{equation}
where we used that $D_\mu\Fcal^{\mu \nu}=0$ since $\Acal_\mu$ is an instanton on $X$.

\medskip

\noindent {\bf Projection on $\Mcal$}.  For $t\in\R$ varying, the connection $\Acal=\Acal (\phi^\a(t), x^\mu)$ on
the bundle $E\to \{t\}\times X$ defines a map
\begin{equation}\label{3.9}
\phi : \R\to\Mcal\with \phi(t) = \{\phi^\a(t)\}\ ,
\end{equation}
where $\phi^\a$ with $\a =1,...,$dim$_{\R}\Mcal$ are local coordinates on $\Mcal$. This map is not free - it is
constrained by the equations (\ref{3.7})-(\ref{3.8}). Since $\Acal$ belongs to the solution space $\Ncal$ of the
instanton equation (\ref{2.4}), its derivative $\partial_0\Acal$ is a solution of the linearized form of (\ref{2.4})
around $\Acal$, i.e. $\partial_0\Acal$ belongs to the vector space $T_{\Acal}\Ncal$. Using (\ref{2.7}), one can decompose
$\partial_0\Acal_\mu$ into two parts,
\begin{equation}\label{3.10}
T_{\Acal}\Ncal =\pi^*T_{[\Acal]}\Mcal\oplus T_{\Acal}\Gcal\quad\Leftrightarrow\quad\partial_0\Acal_\mu
=(\partial_0\phi^\a)\xi_{\a\mu} + D_\mu\eps_0\ ,
\end{equation}
where $\xi_\a=\xi_{\a\mu}\diff x^\mu$ is a local basis of vector fields on $\Mcal$ and $\eps_0$ is an su$(k)$-valued
gauge parameter which is determined by the gauge-fixing equations
\begin{equation}\label{3.11}
g^{\mu\nu}D_\mu\xi_{\a\nu}=0
\end{equation}
and therefore from (\ref{3.10}) and (\ref{3.11}) we get
\begin{equation}\label{3.12}
g^{\mu\nu}D_\mu\partial_0\Acal_\nu=g^{\mu\nu}D_\mu D_\nu\eps_0\ .
\end{equation}
Note that
\begin{equation}\label{3.13}
\Fcal_{\nu 0}= D_\nu\Acal_0-D_0\Acal_\nu\ .
\end{equation}
Let us fix the gauge of the Yang-Mills fields on $\R\times X$ by choosing
\begin{equation}\label{3.14}
\Acal_0:=\eps_0\ .
\end{equation}
Then from (\ref{3.10}) we obtain
\begin{equation}\label{3.15}
\Fcal_{\nu 0}= - \dot\phi^\a\xi_{\a\nu}\ ,
\end{equation}
where we denoted by dot the derivative with respect to time $t$. From (\ref{3.11}) and
(\ref{3.15}) we see that the equations
(\ref{3.7}) are satisfied. Furthermore, since
\begin{equation}\label{3.16}
\partial_0\Acal_\mu=  \dot\phi^\a\frac{\partial\Acal_\mu}{\partial\phi^\a}\ ,
\end{equation}
we get from (\ref{3.12}) that
\begin{equation}\label{3.17}
\Acal_0=\eps_0=\dot\phi^\a\eps_\a\ ,
\end{equation}
where the gauge parameters $\eps_\a$ can be obtained as solutions of the equations
\begin{equation}\label{3.18}
g^{\mu\nu}D_\mu D_\nu\eps_\a= g^{\mu\nu}D_\mu \frac{\partial\Acal_\nu}{\partial\phi^\a}\ ,
\end{equation}
which follow from (\ref{3.12}),(\ref{3.16}) and (\ref{3.17}). Notice that $\Fcal_{0\mu}$,
given in (\ref{3.15}), is the projection of $\partial_0\Acal_\mu$ from $T_{\Acal}\Ncal$ to $T_{[\Acal]}\Mcal$ (cf.~\cite{8}):
\begin{equation}\label{3.19}
\pi_*\partial_0\Acal_\mu = \Fcal_{0\mu}=\dot\phi^\a\xi_{\a\mu}\ .
\end{equation}

\medskip

\noindent {\bf Geodesics}. Although the evolution of the gauge fields does not exactly follow a trajectory $\phi^\a(t)$
in the set of exact static solutions (moduli space $\Mcal$ of instantons on $X$ in our case), it does a good
approximation. Following \cite{20}, we will show that in the adiabatic limit $\ve\to 0$ the approximation becomes exact
and $\phi (t)$
is a geodesic motion on $\Mcal$. To show this, we substitute (\ref{3.15}) in the remaining unsolved equations (\ref{3.8})
and obtain
\begin{equation}\label{3.20}
g^{\mu\nu}\frac{\diff}{\diff t}\left (\dot\phi^\b\xi_{\b\nu}\right ) = g^{\mu\nu}\dot\phi^\b[\xi_{\b\nu},\eps_0]\ .
\end{equation}
Now let us multiply these equations on $\dot\phi^\a\xi_{\a\mu}$, take trace $\tr$ over su$(k)$ and integrate over $X$.
We get the equations\footnote{The right hand side of (\ref{3.20}) vanishes since $ g^{\mu\nu}\dot\phi^\a\dot\p^\b\tr([\xi_{\a\mu},\xi_{\b\nu}],
\eps_0)\equiv 0$ due to converting symmetric and antisymmetric in $(\a\b)$ parts.}
\begin{equation}\label{3.21}
\frac{\diff}{\diff t}\left (G_{\a\b}\,\dot\phi^\a\,\dot\phi^\b\right ) =0
\end{equation}
on the moduli space $\Mcal$. In deriving (\ref{3.21}) we identify $t$ with the affine parameter $s$ entering
in definition of the metric
\begin{equation}\label{3.22}
\diff s^2= G_{\a\b}\,\diff\phi^\a\,\diff\phi^\b
\end{equation}
on $\Mcal$, where the metric components $G_{\a\b}$ were introduced in (\ref{2.3}):
\begin{equation}\label{3.23}
G_{\a\b}=\int\,g^{\mu\nu}\, \tr (\xi_{\a\mu}\wedge *\xi_{\b\nu})
\end{equation}
with the Hodge operator $*$ on $X$.

Equation (\ref{3.21}) defines geodesics on $\Mcal$. To see them in more standard form, with Christoffel symbols
\begin{equation}\label{3.24}
\Gamma^{\a}_{\b\ga}=G^{\a\la}\left(\frac{\partial}{\partial\phi^\ga}\, G_{\b\la}+
\frac{\partial}{\partial\phi^\b}\, G_{\a\la}- \frac{\partial}{\partial\phi^\la}\, G_{\a\b}\right)\ ,
\end{equation}
we consider the action functional
\begin{equation}\label{3.25}
\tilde{S} = \int \diff t\sqrt{G_{\a\b}\dot\phi^\a\,\dot\phi^\b}\ .
\end{equation}
The Euler-Lagrange equations for (\ref{3.25}) are
\begin{equation}\label{3.26}
\ddot\phi^\a + \Gamma^\a_{\b\ga}\dot\phi^\b\dot\phi^\ga -
\dot\phi^\a\frac{\diff}{\diff t}\ln(G_{\b\ga}\dot\phi^\b\dot\phi^\ga) =0\quad\stackrel{(3.21)}{\Longrightarrow}\quad
\ddot\phi^\a+\Gamma^{\a}_{\b\ga}\,\dot\phi^\b\,\dot\phi^\ga =0
\end{equation}
In other words, (\ref{3.20})-(\ref{3.21}) yield equations (\ref{3.26}) of geodesics on the moduli
space $\Mcal$ of instantons on $X$. This also reflects the well-known (classical) equivalence of the action functional
(\ref{3.25}) and the functional
\begin{equation}\label{3.27}
{S} = \int \diff t\,{G_{\a\b}\dot\phi^\a\,\dot\phi^\b}
\end{equation}
for which (\ref{3.26}) are the Euler-Lagrange equations. Note that (\ref{3.27}) is the effective action for the
standard Yang-Mills action functional on $\R\times X$ in the limit $\ve\to 0$. It stems from the term
\begin{equation}\label{3.28}
\int_M d\,\mbox{vol}\,  \tr (\Fcal_{0\mu}\Fcal^{0\mu})
\end{equation}

Finally, note that the pair $(\Acal_0 (\phi (t)), \Acal_\mu (\phi (t), x))$ can be understood as a connection on $\R\times X$
which obeys part of the Yang-Mills equations and in the neighbourhood of $(\Acal_0, \Acal_\mu )$ there is a solution of the
full Yang-Mills equations with $\ve\ne 0$ at least for $\ve$ sufficiently small (cf.~\cite{8, 11, 7}). This follows from the
implicit function theorem and means also the bijectivity of moduli space of the time-dependent solutions for $\ve =0$ and
small $\ve\ne 0$ (cf.~\cite{11, 7}).

\medskip

\noindent{\bf Monopoles and vortices}. It is well known that instanton equations on $X^d$ can be reduced
to monopole equations on a submanifold $X^{d-1}$ in $X^d$ and similarly (generalized) vortex equations can
be obtained by a reduction on a submanifold $X^{d-n}$ with $n\ge 2$. That is why we will be brief and mention
only some examples.

Canonical example is given by the case $d=4$. Considering $X^4=\R^4$ and imposing translation invariance with
respect to the fourth coordinate $x^4$ on $\R^4$ one sees that anti-self-dual Yang-Mills equations on $\R^4$ are reduced
to the Yang-Mills-Higgs Bogomolny equations on $\R^3$ describing non-abelian monopoles~\cite{1,2, 5,6}. Then our consideration
produces geodesics on the monopole moduli space reproducing Manton's result~\cite{8}. In principle, the same can
be done for $d>4$. For example, monopoles on $G_2$-holonomy manifolds $X^7$ can be obtained from Spin(7)-instantons on $X^8$
as in~\cite{21}.

Similarly, as was mentioned in the Introduction, the anti-self-dual Yang-Mills equations on the manifold
$X^4=\Sigma_2\times S^2$ are reduced by imposing SO(3)-symmetry to vortex equations on a Riemannian 2-manifold $\Sigma_2$
(see e.g.~\cite{2,3} and references therein). In other words, vortices on $\Sigma_2$ can be considered as
SO(3)-symmetric instantons on $\Sigma_2\times S^2$. Then the adiabatic approach to the Yang-Mills equations on
$\R\times \Sigma_2\times S^2$ yields to geodesics on vortex moduli space. The same reduction from instantons to
vortices can be done for $d>4$
(see e.g.~\cite{22, 23}) for $X^d=X^{2p}\times X^{2q}$ with K\"ahler manifolds $X^{2p}$ and $X^{2q}$. Then one obtains
generalized vortex equations (see e.g.~\cite{22, 23}) on $X^{2p}$ and the adiabatic approach will describe slowly
moving vortices via geodesics on moduli space of vortices on $X^{2p}$ or symmetric instantons on $X^{2p}\times
X^{2q}$.

\bigskip

\noindent
{\bf Acknowledgements}

\noindent
This work was partially supported by the Heisenberg-Landau Program.


\end{document}